\documentclass[pdflatex,sn-mathphys-num]{sn-jnl}

\usepackage{graphicx}%
\usepackage{multirow}%
\usepackage{amsmath,amssymb,amsfonts}%
\usepackage{amsthm}%
\usepackage{mathrsfs}%
\usepackage[title]{appendix}%
\usepackage{xcolor}%
\usepackage{textcomp}%
\usepackage{natbib}
\usepackage{manyfoot}%
\usepackage{booktabs}%
\usepackage{algorithm}%
\usepackage{algorithmicx}%
\usepackage{algpseudocode}%
\usepackage{listings}%
\usepackage[symbol]{footmisc}%

\theoremstyle{thmstyleone}%
\theoremstyle{thmstyletwo}%

\theoremstyle{thmstylethree}%

\begin{document}

\title[Article Title]{\textit{RoboComm}: A DID-based   scalable and privacy-preserving Robot-to-Robot interaction over state channels \footnote[2]{This paper is currently under peer review with possible modifications. Subject to revision based on reviewers feedback.}
}


\affil[1]{\orgdiv{Scurid Incorporation, Japan}}

\author*[1]{\fnm{Roshan} \sur{Singh}}\email{roshan@scurid.com}

\author[1]{\fnm{Sushant} \sur{Pandey}}\email{sushant.pandey@scurid.com}

\abstract{
In a multi-robot system establishing trust amongst untrusted robots from different organisations while preserving a robot's privacy is a challenge. Recently decentralized technologies such as smart contract and blockchain are being explored\cite{singh2020efficient}\cite{dorigo2024blockchain} for applications in robotics. However, the limited transaction processing and high maintenance cost hinder the widespread adoption of such approaches. Moreover, blockchain transactions be they on public or private/permissioned blockchain are publically readable(in the case of a private/permissioned blockchain, the data is accessible within the network participants) which further fails to preserve the confidentiality of the robot's data and privacy of the robot. 
\\

\hspace{0.5cm} In this work, we propose \textit{RoboComm} a Decentralized Identity(DID) based approach for privacy-preserving interaction between robots. With DID a component of Self-Sovereign Identity(SSI); robots can authenticate each other independently without relying on any third-party service. Verifiable Credentials(VCs) enable private data associated with a robot to be stored within the robot's hardware, unlike existing blockchain based approaches where the data has to be on the blockchain. We improve throughput by allowing message exchange over state channels. Being a blockchain backed solution \textit{RoboComm} provides a trustworthy system without relying on a single party. Moreover, we implement our proposed approach to demonstrate the feasibility of our solution.
}




\maketitle
\section{Introduction}
Robots are now a part of society and are no longer restricted within bounded walls inside manufacturing units. These social robots \cite{bartneck2004design} with human-like visual, perception, and speech cues interact with humans and the open environment with little to no restrictions. A social robot gathers data through interaction with humans, whereas a swarm of homogeneous robots on a mission exchange messages in real time to reach a consensus on critical decision-making. The same applies to robot-to-robot interactions in a multi-robot system\cite{gielis2022critical}. Today, there is a lack of a transparent, accountable, and privacy-preserving framework for data provenance to monitor these interactions. This raises concerns about the confidentiality and privacy of data from an individual robot\cite{ferrer2023if}. The exchanged data can be processed and even sold by the data receiver without the consent of the data producer. 
 \begin{figure}[h!]
    \centering
    \includegraphics[width=110mm]{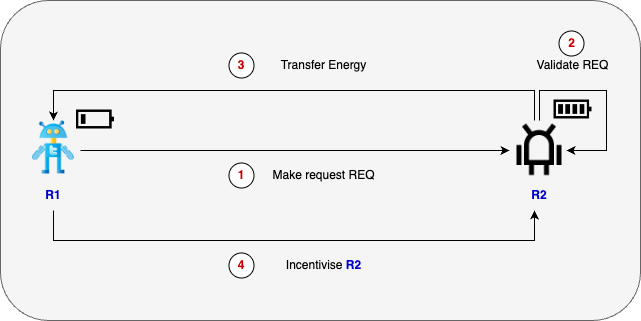}
    \caption{Steps in P2P energy transfer between two robots}
    \label{fig:ett}
\end{figure}
Consider a scenario as shown in Fig. \ref{fig:ett} where a mobile robot $R_{1}$ performing a data processing task(e.g. matrix multiplication) runs low on battery and needs to be charged to complete its task. It finds a stationary idle robot $R_{2}$ with surplus battery backup open to share energy. The $R_{2}$ need to authenticate and verify certain information, such as end of life and background of $R_{1}$ before, $R_{1}$ can charge itself. If  $R_{1}$ discloses its manufacturing information, whereas the manufacturing information contains the manufacturer's name, hardware specification, date of the end of life etc., the robot $R_{1}$ will be disclosing more private information besides the date of the end of life. Moreover, robot authentication is also a challenge. A robot needs to rely on a centralized service for identity management, authentication, and credential issuance. These centralized services have availability issues and can become a central point of failure hampering the robotic operations.  

\section{Motivation}
To address the above concerns, we propose to use W3C Decentralised Identity and Verifiable Credentials (VCs) for data provenance along off-chain communications for data exchange. Robots can generate their DIDs and obtain VCs through designated issuers. Each claim is individually signed by the issuer. This allows the robot to selectively disclose the required information instead of disclosing all information. Data exchange between the robots can take place off the chain through peer-to-peer connections for faster exchange.
Thus the main contributions of this work are as follows:
\begin{enumerate}
    \item Provide robots a SSI with DID-based decentralized authentication and a p2p communication mechanism.
    \item Increase transaction throughput of blockchain-based system with off-chain transactions on the Layer 2 state channels.
    \item Provide a trustworthy shared platform for robots from different organizations to buy and sell services in the ecosystem, thus boosting the autonomous robot economy.
\end{enumerate}

\section{Preliminaries}
In this section, we discuss the technologies and the primitives used for designing our architecture.
\subsection{Channel Network}
A channel network is a Layer 2 scaling solution. In a channel network, two parties, Alice and Bob, create a channel that enables them to keep a large number of transactions "offchain." To put it another way, the ledger is only used when the parties involved in the channel disagree or wish to close the channel. If the parties are not in conflict, they are free to update the balance in the case of a payment channel or the smart contract state in the case of a ledger channel. Individuals will strongly prefer honest behaviour since there is no motivation for disagreement because off-chain transactions can always be handled on a ledger by the individuals concerned. In the best-case scenario, channels greatly lower transaction costs, enable fast payments, and lessen the strain on the ledger when both parties to the payment channel operate honourably and off-chain transactions never appear on the ledger before the channel is closed.

\begin{figure}
    \centering
    \includegraphics[width=130mm]{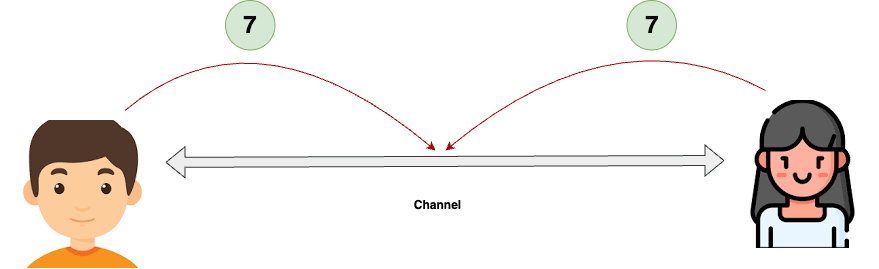}
    \caption{A Payment channel between Bob(left) and Alice(right)}
    \label{fig:arch}
\end{figure}

 In case of any misbehaviour from a participant or the participant simply disappears from the channel in between the interaction, the other participant can always recover his funds without any loss by producing the latest signed transaction from the misbehaving participant. Also, the misbehaving party cannot obtain extra tokens as each transaction needs to be digitally signed by both participants. The Fig. \ref{fig:arch} shows a payment channel established between Alice and Bob. Both put 7 coins/tokens each onto the channel. These tokens get locked on the channel and cannot be spent until the channel is closed. 

 State channels are no different and work similar to the payment channels. However, instead of only transferring financial information about one's account balance, the state channels also facilitate sharing account state information. In case of smart contract the account can maintain arbitrary values, such as the battery level and credit score of a robot with a designated address on the blockchain.

\subsection{Decentralised Identity(DID) and DID Communication}
\subsubsection{Decentralised Identity}
A Decentralized Identifier (DID) is a globally unique, cryptographically verifiable identifier that does not require a centralized registry or authority for its issuance or management. DIDs are a core component of decentralized identity systems and are defined by the W3C DID Specification. They are designed to enable self-sovereign identity—a model in which individuals or entities fully control their digital identifiers and associated data.
A typical DID structure consists of the three mandatory components:- 
\begin{center}
\textbf{$<did:method:method-specifier>$}
\end{center}
Eg. 
\begin{center}
\textbf{$<did:robo:0xAb8483F64d9C6d1EcF9b849Ae677dD3315835cb2>$}
\end{center}
where, the meaning of the respective components is as follows:
\begin{enumerate}
    \item $did$: The scheme indicating that it's a decentralized identifier.
    \item $method$ : Specifies the DID method, which defines how the identifier is created, resolved, and managed on a specific blockchain or distributed system.
    \item $method-specifier$ : A unique string generated according to the method’s rules, often linked to a public key or hash.
\end{enumerate}
In addition, a DID document serves as a metadata container that describes the cryptographic and service-related capabilities associated with a specific decentralized identity (DID). It typically includes one or more public keys used for cryptographic operations such as authentication, authorisation, and data integrity verification. These keys enable secure interactions between entities without relying on centralized intermediaries. In addition to public keys, the DID document may define one or more service endpoints, URLs, or network addresses that facilitate interactions with services controlled by the DID subject or its controller. These endpoints can support use cases such as credential issuance, verification services, messaging, or decentralized data exchange. The structure and semantics of the DID document conform to the W3C DID specification, ensuring interoperability across different platforms and implementations.
\subsubsection{DID Communication(DIDComm)}
DIDComm provides a secure and private communication methodology built atop the decentralized design of DIDs. It is transport agnostic and supports both simplex and duplex; works offline; doesn’t assume client-server or synchronous or real-time; allows paired or n-wise or public broadcast usage\cite{didcomm}. Unlike today's communication methodologies, which mostly rely on key registries, identity providers, certificate authorities, browser or app vendors, or similar centralizations, DID-based communication allows decentralised, peer-to-peer, verifiable communication without relying on any centralised service or authority.
\subsection{Verifiable Credentials (VC)}
A verifiable credential (VC) is a tamper-evident, cryptographically secure digital assertion made by an issuer about a subject. It includes three key components: "metadata," which provides contextual details (e.g., issuer, issuance date, and public key references); "claims," which are statements about the subject (e.g., qualifications or certifications); and "proof," a cryptographic signature ensuring authenticity and integrity. VCs enable trustless, portable, and privacy-preserving information exchange in decentralized identity systems, supporting "self-sovereign identity" (SSI). They allow secure, offline verification without relying on centralized authorities, while also enabling features like selective disclosure and zero-knowledge proofs for enhanced privacy.

In recent years, there have been a few works done in order to integrate DID and VCs into the Internet of Things space. In \cite{dayaratne2024ssi4iot} authors provide a  comprehensive taxonomy and usage of VCs in the IoT context with respect to their validity period, trust and interoperability level, and
scope of usage. The life-cycle management of VCs as well as various optimisation techniques for realising SSI in IoT environments are
also addressed in great detail.
\subsection{Sovereign Robots}
A sovereign robot is equipped with a Self-Sovereign Identity (SSI)—a decentralised, verifiable digital identity that operates independently of centralized or federated identity providers. SSI grants the robot full autonomy over the creation, control, and management of its identity, ensuring maximum portability, resilience, and security. Leveraging Decentralized Identifiers (DIDs) and Verifiable Credentials (VCs), the robot can establish cryptographically verifiable claims and engage in trusted, peer-to-peer communications with other entities in a decentralized ecosystem.

\subsection{Layer 1 Blockchain}
A layer 1 blockchain such as Bitcoin  and  Ethereum is a decentralised network of nodes connected with a peer-to-peer network. These nodes maintain the blockchain ledger with its own consensus mechanism and doesn't depend on any third entity for this purpose. A Layer 1 blockchain can also have support for smart contracts, which allows custom codes to be written in a high-level language that can be deployed and executed on the blockchain.
\section{Proposed Architecture}
In this section, we describe our proposed state channel-based architecture for robot-to-robot interactions. Figure \ref{fig:roboframework} shows the proposed architecture. The architecture consists of two layers, Layer 1 and Layer 2. Layer 1 is a blockchain running its own consensus mechanism, such as Proof-of-Work, Proof-of-Stake, and Byzantine Fault Tolerance (BFT). The Layer 1 blockchain also supports smart contracts, which allow business logic to be implemented as code that can be executed on the blockchain. Organisations deploying robots in the ecosystem can come together to setup a private/permissioned blockchain. Existing public Layer 1 blockchains with smart contract support such as Ethereum and Cardano can also be used for better transparency and decentralisation with a compromise in customisation.

Meanwhile the Layer 2 consists of multiple robots of different types, characteristics and manufacturers from different organisations. In a practical and real life scenario, two robots from two different organisations with no trust in each other can come forward to initiate a secure and verifiable interaction by verifying each other's information from the Layer 1 blockchain. A robot from an organisation trusts its peer robots from the same organisation; these robots communicate with each other in a P2P manner to access blockchain information. However, due to storage, communication and energy constraints on a robot not all robots maintain an upto date information about the blockchain ledger. A few robots from an organisation that have sufficient storage and power backup called a full node are entrusted to maintain up-to-date information on the blockchain ledger. These full nodes are queried by its organisation peers for verifying DIDs of robots from other organisations. 

A set of well-known issuers from different organisations are entrusted to provide VCs to robots. The information regarding the issuance of a credential is kept on the blockchain. A VC is digitally signed by an issuer, which can be independently verified by any robot without any trusted entity. An organisation provides a VC to the robot of its own organisation. The DIDs of the issuers are made available on the blockchain through a smart contract, which a robot can refer to. 

\begin{figure}
    \centering
    \includegraphics[width=130mm]{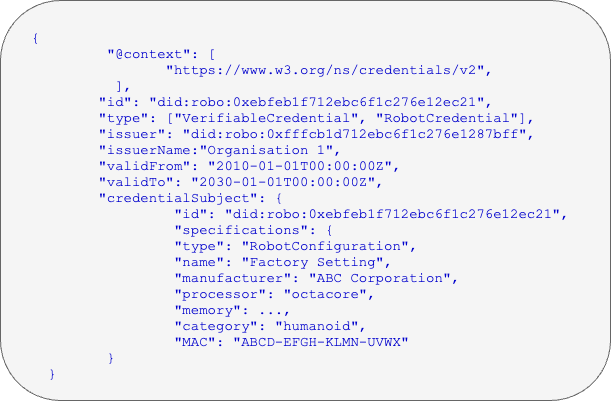}
    \caption{An example of a Verifiable Credential of a robot}
    \label{fig:robovc}
\end{figure}

\begin{figure}
    \centering
    \includegraphics[width=130mm]{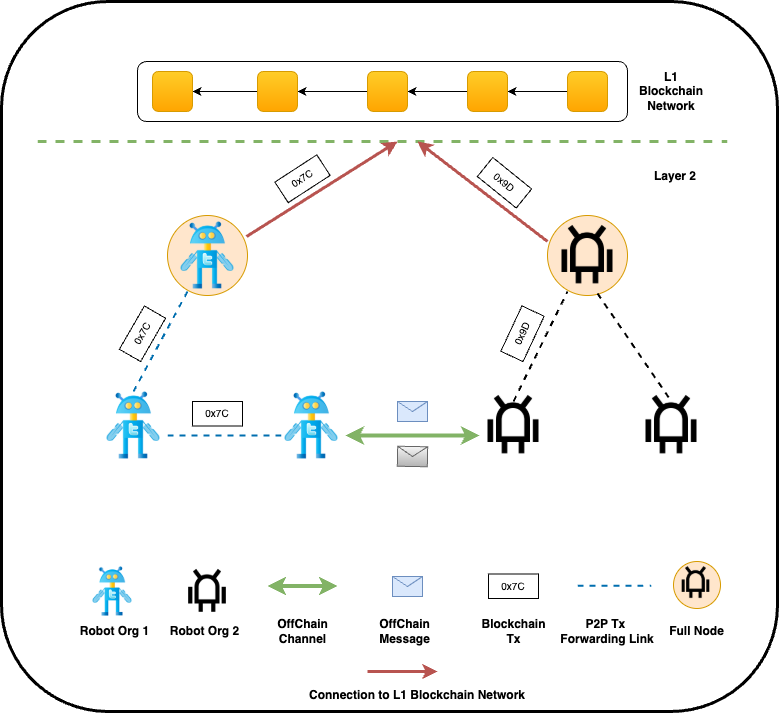}
    \caption{Architecture of RoboComm}
    \label{fig:roboframework}
\end{figure}

However, the mapping of an issuer to an organisation is not made available on the blockchain in order to preserve the privacy of VCs issued to a robot. If the mapping is available, then an adversary can perform a linkage attack between the VCs issued by an issuer, which can compromise a robot's privacy. The VC contains the DID information of the robot along with other specifications. The VC acts as a certificate here, whereas the issuer is the certificate authority. The structure of a VC issued to a robot is shown in Fig. \ref{fig:robovc}.

A robot receiving a VC from a peer robot verifies the certificate's authenticity. The robot then checks with the blockchain the current status of the robot's energy status and credit score with the obtained DID from the robot's VC. Once both the robots are satisfied by the information available on the blockchain, the robots can start off-chain communications. They exchange off-chain transactions digitally signed by the private key associated with the address of its DID. A robot receiving an off-chain transaction verifies the signature and updates its local state. Any one of the robots can initiate a state closure by submitting an on-chain transaction on the Layer 1 blockchain.
\subsection{Use Case: Peer-to-peer energy transfer in robotics}
Robots, such as a delivery robot used for doorstep delivery purposes, need to move throughout the city, navigating paths, avoiding living and non-living obstacles. Because of the need to be mobile, it cannot be hooked up to a regular direct power supply during its operation. This requires a battery-based rechargeable solution. In recent years, these service robots have come equipped with rechargeable batteries. However, during long operations, it is evident that they may run out of power and need to do a recharge. It is not economical to establish a robot charging station in every corner of the city. Again, adding a backup battery is also not a feasible solution, as it is going to increase the weight of the robot and will result in higher power consumption, which means the solution is not viable.

Peer-to-peer energy transfer among robots seems to be a viable solution here. In P2P energy transfer, a robot with a low battery gets its battery recharged by another robot in its vicinity with surplus battery backup\cite{sarin2020code}. This energy transfer occurs mainly wirelessly. Wireless energy transfer (WET) in robotics is a new paradigm in which energy is wirelessly transferred between robots. In recent years, some work has explored WET in robotics\cite{guo2021joint}\cite{WPT}\cite{han2024untethered}\cite{krestovnikov2021development}. Robots can reach a consensus on the energy transfer procedure and initiate the transfer of energy.

However, ensuring proper and fair energy transfer between robots in an unmonitored environment is a challenge. Robots can belong to different organizations and thus their benign behaviour cannot always be granted. Also, because they can come from multiple organizations a centralized solution can't be established. Moreover, being operating in an open environment, they are prone to hacks.

Decentralized solutions such as blockchain can be a viable platform for developing energy transfer solutions for these independent robots\cite{zhou2023decentralized}. The robots can get their VC with a DID registered on the blockchain, enabling it to perform authentication without any third party. However, due to the low throughput of the existing Layer 1 blockchains, it is not an ideal choice to develop the solution. Layer 2 technologies such as state channels can help here. The robots can initiate the energy transfer deal over the layer 1 blockchain with the help of smart contracts.

Once the consensus on the deal is reached the energy transfer starts for each unit of energy transfered from the source to the destination. The source robot signs an off chain transaction and sends it to the destination robot. The transaction contains the information such as the total number of energy units transferred from source to destination, along with other necessary information.

Similarly, the destination robot also releases an off-chain transaction from destination to source with the total unit of credit score transferred to the source in exchange for the energy transfer. A successful transfer of credit score from source to destination results in an increment of credit score to the destination and a decrease in the source's credit score. These earned credit scores can then be used by the robot to avail discounts on services such as maintenance and upgrades, thus enabling a thriving robot economic ecosystem. Extending the concept of micropayments\cite{mercan2021cryptocurrency} from the cryptocurrencies world, we introduce a pay-per-unit model in our energy trade where the seller pays in terms of credit score for each unit of energy received from the buyer.

The transfer of energy in an energy trade between two robots is performed in four phases:-
\begin{enumerate}
    \item \textbf{Robot Discovery} :-
    
    This is the first phase of the energy transfer. In this phase the prospective robot with sufficient energy is identified. 
    \begin{enumerate}
        \item A robot running low on battery broadcasts beacon messages with its DID and the VC issued to it by an issuer. A robot with sufficient energy replies with an affirmative message to start energy transfer. 
 
        \item The robots are identified with the help of a DID, which can be verified on the blockchain. The DID is used for robot identification and for communication. Robots receiving a DID from a peer verify the authenticity and credibility of the broadcasting/sender robot by polling the blockchain for information against the DID. A robot considers the peer robot for the further steps if the robot DID is not revoked and is active. 
    \end{enumerate}
    \item \textbf{Robot Selection}:-
    \begin{enumerate}
        \item There can be multiple robots in the vicinity willing to participate in energy transfer. In this case, the buyer robot checks with the blockchain the credit score of each robot and selects the robot with the highest credit score. The robot selection procedure can be further improved by incorporating a much more complex algorithm; developing such an algorithm is out of scope of this work.
        
    \end{enumerate}
    \begin{figure}[]
    \begin{center}
    \includegraphics[width=130mm]{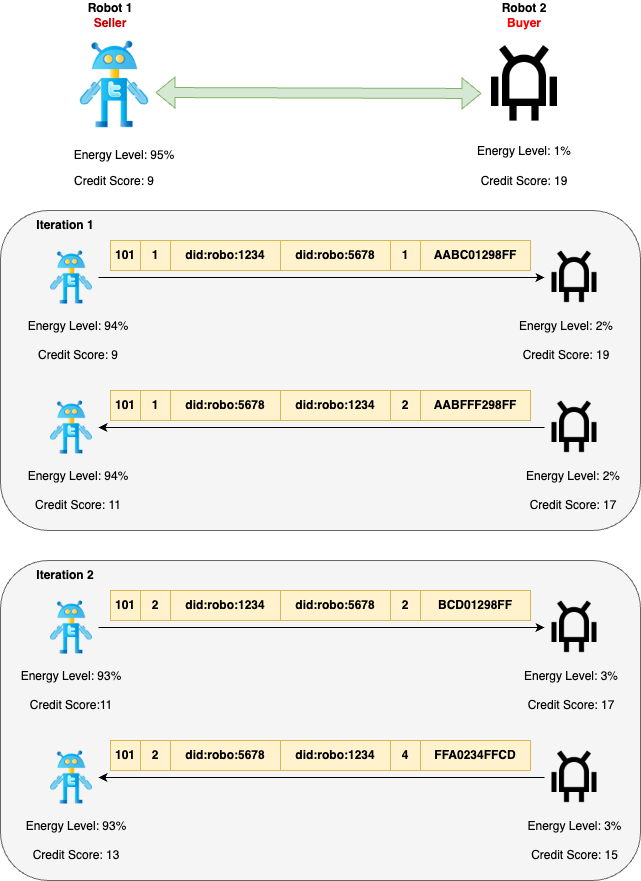}
    \label{fig:seqdia}
\caption{Off-chain interaction between two robots during energy transfer}
\end{center}
\end{figure}

    \item \textbf{Channel Establishment and Energy Transfer}:-
    This phase is concerned with the creation of a ledger channel or state channel between the buyer and seller robots. The buyer robot is the one willing to purchase energy units, whereas the seller robot is the one willing to sell energy units in exchange for credit points. 
    For each successful unit of energy transferred, an off-chain transaction is executed from the seller to the buyer. In response to this, the buyer also executes an off-chain transaction, which updates the credit score of the seller. Fig. 3 shows a sequence diagram for a sample trade between two robots.
\begin{enumerate}
    \item \textit{Channel Opening}: Both the buyer and seller robot execute on chain transactions to open a channel. The channel establishment is controlled by a function deployed as a smart contract on the blockchain. For a channel to be opened, both parties need to provide confirmation by executing transactions respectively.
    \item \textit{Energy Transfer and Credit Score Update}:         The energy transfer phase deals with the transfer of energy from the seller robot to the buyer robot. The off-chain exchange of energy starts. For each unit of energy received from a participating robot in the exchange. The receiving robot needs to send an off-chain transaction to the sender node, with the respective update in the credit score.
    \item \textit{Channel Closing}: This is done by closing the state channel, which triggers the final settlement of the transactions on the blockchain. This final settlement is done by broadcasting the last transaction to the blockchain, which records the final state of the channel, updating the energy levels and credit score of both the seller robot and buyer robot.
\end{enumerate}
\end{enumerate}
The diagram 4 shows the sequence of events performed by the involved parties during the energy exchange. The energy exchange occurs over several iterations. For each iteration a unit of energy is transferred in exchange for some credit score. 
    \begin{figure}[]
    \begin{center}
\includegraphics[width=90mm]{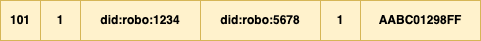}
    \label{fig:txs}

\caption{An off-chain transaction payload}
\end{center}
\end{figure}

Consider the iteration 1 from the diagram where Robot 1 exchanges one unit of energy with Robot 2 for 2 points of credit score. The state updates are made via the offchain transactions. The structure of the offchain transaction is as shown in Fig. 5, which can be broken down as below:-
\begin{enumerate}
    \item \textit{Exchange ID}: This is a unique id for an exchange.
    \item \textit{Iteration Number}: A whole number denoting the iteration in the energy exchange.
    \item \textit{Sender Address}: A blockchain address of the sender robot. The sender robot is the robot signing the transaction.
    \item \textit{Receiver Address}: A blockchain address of the receiver robot. The receiver robot is the robot for which the transaction is intended.
    \item \textit{Value}: A data denoting either the units of energy or a credit score.
    \item \textit{Signature}: The digital signature of the sender robot.
\end{enumerate}

\subsection{Dispute Resolution: How to handle cheating robots?}
Disputes may arise at any stage of the robot-to-robot interaction over the state channel. Some disputes can arise intentionally where a misbehaving robot can attempt to cheat by presenting an outdated state or falsified transactions. However, there are also scenarios where the dispute may arise unintentionally from natural reasons such as communication delays and timeouts where the channel is not closed properly or the participants cannot reach an agreement on the final state. There exist several techniques in the literature that can be used for dispute resolution under such scenarios, such as timelocks and cryptographic commitments. We handle the dispute resolution in our approach as follows:

\begin{enumerate}
    \item \textit{When the dispute occurs because of a cheating robot}: If a cheating robot $R_1$ tries to close the state channel unilaterally with a state that is not agreed upon by both, the robot $R_2$ can initiate a dispute by revealing the most recent valid state. A challenge period is a time window during which one party can contest the closure or the state submitted by the other party. During the challenge period, the disputed state can be reviewed, and the honest robot can reclaim its rightful share by presenting cryptographic evidence or other proofs of the legitimate state. Moreover, punishment mechanisms can also be introduced to punish misbehaving robots in order to encourage honest behaviour.

    \item \textit{When a dispute occurs due to natural reasons}: If the peer robot goes offline or becomes unavailable, for reasons such as hardware or communication failure, the other robot can wait for a threshold $\Delta$ interval and can submit the last signed transaction from the offline robot and on the blockchain if the peer robot doesn't come online.
\end{enumerate}

    In the worst-case scenario, the seller robot might incur a loss of a unit of energy. This will happen when the buyer robot does not release an off-chain transaction, incrementing the credit score of the seller for the extra unit of energy received from the last off-chain transaction executed by the seller. This loss of the seller can be minimized by considering the smallest microunit possible for the energy to be traded/transferred. Moreover, an honest seller can be cheated by a misbehaving buyer once. The seller can maintain a local list of misbehaving robots based on past interaction experience and deny providing services to them later on. Here, we assume that the robot hardware, such as the energy meter, is fault-free and has a negligible error margin.

\subsection{Incentives and Punishments}
Robots are incentivized for their honest behaviour with an increase in their credit score. This happens when the off-chain channel is closed with mutual consent, or in scenario 4.2.2. At the same time, cheating and misbehaving robots are punished with a decrease in their credit score when they attempt to close the channel with an outdated transaction for personal gain.
The deduction in the credit score of the cheating robot occurs as and when a fraud-proof is submitted on the blockchain by the other robot in the channel. Both increase and decrease logic is an integral part of the Energy Trade smart contract.

\section{Implementation}
In this section, we discuss the implementation of our proposed solution in detail.

Our proposed solution makes use of the following technologies:
\begin{enumerate}
    \item \textit{Consensys Quorum}\cite{consensys}: The Consensys Quorum (GoQuorum) is an open-sourced, enterprise-focused blockchain that acts as the layer 1 blockchain as shown in the architecture diagram. It is an EVM-based machine and consists of all the supported functionalities like the Ethereum blockchain; the only difference is that it does not have the concept of gas and gas prices. Quorum also has support for private transactions and private contracts through public/private state separation.
    
    \item \textit{libp2p}\cite{libp2p}: libp2p is an open-source networking library with a modular network stack used to build peer-to-peer (P2P) applications. It is the underlying networking layer that powers many decentralized projects, including IPFS (InterPlanetary File System). libp2p provides a flexible set of tools for creating P2P networks, allowing developers to build secure and scalable distributed applications.

    \item \textit{Solidity}\cite{solidity}: It is a high-level, statically-typed programming language designed for writing smart contracts that run on the Ethereum Virtual Machine (EVM). Smart contracts are self-executing contracts with the terms of the agreement directly written into code. Solidity allows developers to create decentralized applications (dApps) on the Ethereum blockchain by defining the rules and logic of the contract, which is then executed on the Ethereum network.
\end{enumerate}

We implement the robot clients in Typescript. The p2p communication between the clients is facilitated by the libp2p protocol. The Quorum blockchain is used as layer 1, with the smart contracts written in Solidity. 

Each robot in the system is globally identified with a DID. The P2P communication between the two robots is established with libp2p multi-address. A multiaddress or multiaddr, is a convention for encoding multiple layers of addressing information into a single “future-proof” path structure. It defines human-readable and machine-optimized encodings of common transport and overlay protocols and allows many layers of addressing to be combined and used together. A multiaddr looks like as follows:
\\ \\
{\footnotesize \textit{/ip4/127.0.0.1/tcp/10333/p2p/12D3KooWDTptSsELd2qug4rcXcFoRZNd7mT6NT3cPnZAJ7mH8ydR}} \\

The mapping between a robot DID and its multiaddress is stored in the smart contract on the blockchain. A robot receiving a DID from its peer obtains the peer robot libp2p multiaddr by resolving the peer DID with the blockchain. Upon successful resolution of a robot DID, the querying robot receives a DID document as shown in Fig. \ref{fig:robodoc}, which contains information such as multiaddr, authentication type, controller, etc. 

We deploy four smart contracts:-
\begin{enumerate}
    \item \textit{Registry Contract}: It keeps track of robot DID and its mapping to the robot multiaddress. It also stores metadata about the robot DID, such as its creation time, and status, such as whether the robot DID is active or revoked.

    \item \textit{Issuer Contract}: It keeps track of the DIDs of the well-known issuers, which are authorized to issue verifiable credentials (VCs) to the robots.
    
    \item \textit{Energy Trade Contract}: The trade contract deals with the business logic of the energy trade.

    \item \textit{OnChain Verifier Contract}: It is responsible for verification of off-chain transactions on the blockchain for settlement and channel closure. The smart contract verifies the signature and validates the transaction order.
\end{enumerate}

\begin{figure}[h]
\begin{center}
\includegraphics[width=130mm]{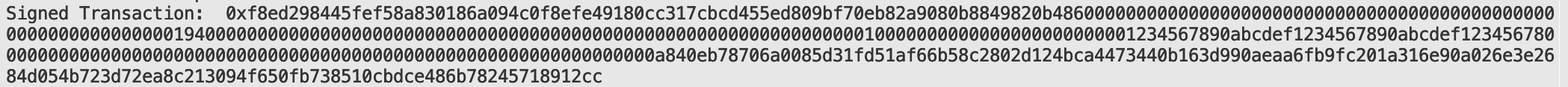}
\caption{Digitally signed off-chain transaction}
\label{fig:signtx}
\end{center}
\end{figure}

\begin{figure}[h]
\begin{center}
\includegraphics[width=130mm]{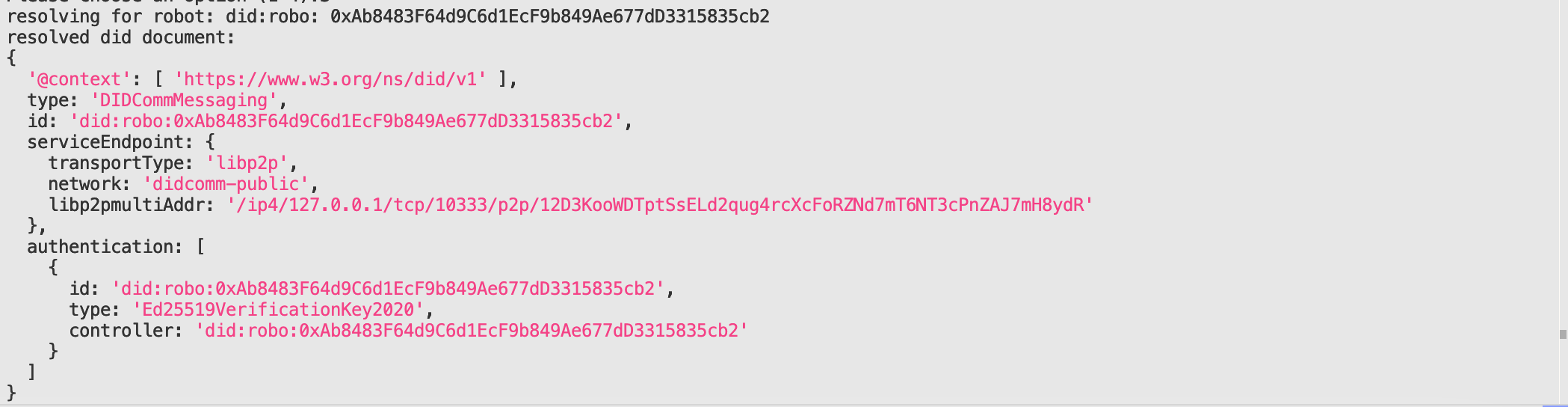}
\caption{DID Document of a robot}
\label{fig:robodoc}
\end{center}
\end{figure}

Once a robot obtains the multiaddr of its peer, it sends a digitally signed off-chain transaction to its peer. A digitally signed off-chain transaction looks like the one shown in Fig. \ref{fig:signtx}. The blockchain rejects an outdated transaction submitted by a misbehaving robot in order to obtain an increased credit score or energy units.

\section{Experiment}
We measure the effectiveness of \textit{RoboComm}, by applying it to robots performing a collaborative payload transport task. We rely on Agent Based Modeling(ABM) for the experiment. ABM is an effective approach for modeling the behavior of swarm robots\cite{oprea2018agent}\cite{gumahad2025simulating}. Using the Mesa\cite{mesa} agent-based modeling framework, a swarm of robots is deployed on a structured grid and tasked with delivering payloads to clustered goal zones modeled via a Gaussian Mixture distribution. Each robot in the swarm is assigned a delivery goal. Details about the setting of parameters for the modeling can be found in Table \ref{table:t1}.

\begin{figure}
    \centering
    \includegraphics[width=130mm, height= 65mm]{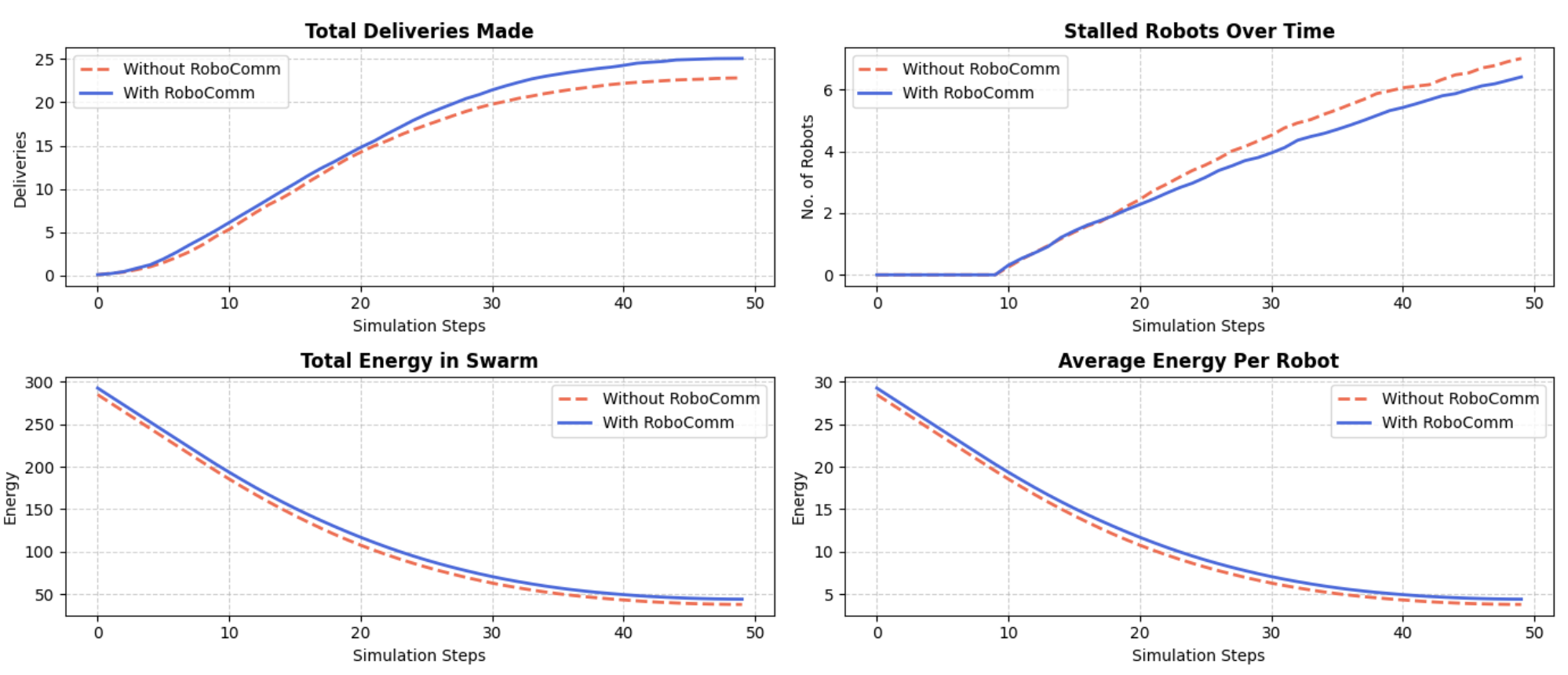}
    \caption{Agent based modeling results of swarm performance with and without \textit{RoboComm} for collaborative payload transport task}
    \label{fig:plots}
\end{figure}

Two configurations are compared: a baseline swarm operating without energy sharing and a \textit{RoboComm}-enabled swarm where low-energy robots can receive energy transfers from nearby idle robots. Performance is assessed on 100 independent runs of 50 simulation steps, with metrics including total deliveries, stalled robots, swarm energy dynamics, and average robot energy as shown in Fig. \ref{fig:plots}. 
\begin{center}
\begin{table}[h]
\begin{tabular}{ |p{4cm}|p{4cm}|  }
\hline
\multicolumn{2}{|c|}{Parameter Setting} \\
\hline
Parameter & Value/Details \\
\hline
No. of Robots & 10  \\
Grid Size  & 20 X 20   \\
Initial Energy & [10, 50] Units \\
Delivery Goal & 5 deliveries/robot \\
Energy Consumption & 1 unit/step  \\
Transfer Trigger & Receiver Energy $\leq$ 2 Units   \\
Delivery Zone Placement & 5 clusters, 50 points each, $\sigma$=1.5 \\
Simulation Steps & 50  \\
Runs  & 100   \\
\hline
\end{tabular}
\caption{Experiment parameters}
\label{table:t1}
\end{table}
\end{center}
\section{Results}
\begin{figure}[h]
\begin{center}
\includegraphics[width=75mm]{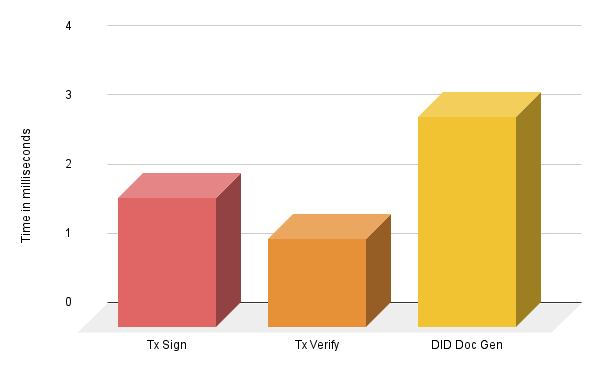}
\caption{Average transaction and DID document generation time}
\label{fig:result}
\end{center}
\end{figure}

We evaluate \textit{RoboComm} performance based on the average transaction time of off-chain state channel operations (TxSign, TxVerify) and DID document generation time (DID Doc Gen). The implementation
is run on a Raspberry Pi 5 with a 2.4 GHz, A-Series CPU and 8 GB of RAM. Moreover, the blockchain network runs on a MacBook Pro laptop with an Apple M2 chip and 8GB of RAM. Fig. \ref{fig:result} shows the graph of time required for signing an off-chain transaction, verifying an off-chain transaction, and DID document generation. The statistics are averaged over 1000 independent runs.

We also measure the size of the robot DID document and the payload of the off-chain transaction as shown in table \ref{table:t2} .

\begin{center}
\begin{table}[h]
\begin{tabular}{ |p{3cm}|p{3cm}|  }
\hline
\multicolumn{2}{|c|}{Size Statistics} \\
\hline
Parameter & Bytes \\
\hline
DID Document & 563  \\
Off Chain Tx.  & 480   \\
\hline
\end{tabular}
\caption{Size of DID Document and Off-Chain tx.}
\label{table:t2}
\end{table}
\end{center}
The preliminary results indicate that \textit{RoboComm} offers the necessary scalability with limited storage overhead. Millisecond range off-chain transactions (signature and verification) over the state channel and the generation time for the robot's DID document (1.873 ms., 1.28 ms., and 3.046 ms., respectively), clearly demonstrates the feasibility of the approach for real world deployments. Further, results from the agent based modeling shows that \textit{RoboComm} reduces premature robot failures, balances swarm energy distribution, and thereby increases task completion efficiency compared to the non energy sharing approach.

\section{Discussion}
With the rise of intelligent and autonomous robots, ensuring secure and privacy-preserving interactions among these robots is of utmost importance. Centralised solutions exist, but they fail to scale, are not transparent, verifiable, and lack trust. RoboComm attempts to tackle these challenges by utilizing decentralized technologies; DID, SSI, and blockchain. Now, robots can authenticate each other without relying on a centralized third party. Moreover, it introduces a state channel-based approach for secure peer-to-peer interaction between robots. Incorporating a state-channel approach extensively reduces the load on the layer 1 blockchain by reducing the number of on-chain transactions executed, thus increasing the scalability of the system. Moreover, being decentralized with deals initiated and settled on the blockchain, this approach provides a common trusted platform for robots from different organizations that potentially do not trust each other. This allows for an autonomous, thriving robot economy where robots across organizations can come forward to offer and purchase services such as energy trade and get incentivized. The robots can then use these earned incentives to avail discounts on maintenance services and upgrades.

With the notion of pay-per-unit, the work preserves the interests of both the buyer robot and the seller robot. This minimizes the loss of credit score and energy units from risks associated with getting cheated by one another. The buyer only increases the seller's credit score if and only if it receives the energy unit. At the same time, the seller also transfers a unit of energy if and only if it receives the increment in credit score from the previous iteration.

\section{Related Works}
Privacy and trust management in robotics are sufficiently studied in the literature. A taxonomy of privacy constructs
for privacy-sensitive robotics is provided in \cite{rueben2017taxonomy} whereas \cite{rueben2018themes} provides themes that should comprise privacy-sensitive robotics
research. In \cite{castello2019blockchain} \cite{strobel2018managing} authors introduce a blockchain-based framework for handling misbehaving and byzantine robots in the system. A blockchain-based approach for collaborative decision-making with a lightweight consensus mechanism is proposed in \cite{singh2020efficient}. Moreover, trust management and trustworthiness in robotic systems are studied in\cite{wilson2023trustworthy} \cite{vojnar2024scenario}\cite{zikratov2016dynamics}. Some works in the literature have also taken advantage of blockchain for trust and reputation management, such as \cite{mallikarachchi2022managing}\cite{li2021blockchain}. Although blockchain-based systems are better than centralized systems, they are not scalable enough for practical deployments due to low transaction throughput and resource-intensive consensus algorithms, which are not ideal for low-resource robots such as swarm robots, which have limited processing and storage resources available.

We have not found any work in the literature that uses SSI and DID-based communication to enhance the privacy of robot-to-robot interaction. To our knowledge, this is the first time SSI technology with DID is being used with a state channel for off-chain peer-to-peer communication in robotics.

\bibliography{sn-bibliography}
\end{document}